\documentclass[apjl]{emulateapj}
\usepackage{graphicx}
\usepackage[dvips]{color}
\usepackage{txfonts}
\usepackage{natbib}
\usepackage{times}

\newcommand{\kms}{\ensuremath{\mathrm{km\,s}^{-1}}}
\newcommand{\msol}{\ensuremath{\rm{M}_{\odot}}}
\newcommand{\teff}{\ensuremath{T_{\rm{eff}}}}
\newcommand{\logg}{\ensuremath{\log g}}

\newcommand{\lheh}{\ensuremath{\log N_{\mathrm{He}}/N_{\mathrm{H}}}}

\newcommand{\uHz}{\ensuremath{\mu{\rm{Hz}}}}
\newcommand{\ebv}{{E($B$--$V$)}}
\newcommand{\kep}{{\em Kepler}}
\newcommand{\kepmi}{{\em Kepler Mission}}
\newcommand{\mkep}{\ensuremath{Kp}}

\newcommand{\sigft}{\ensuremath{\sigma{_{\rm{FT}}}}}

\newcommand\sval[3]{\ensuremath{#1}\ensuremath{#2}\,{#3}}
\newcommand\val[2]{\ensuremath{#1}\,{#2}}
 
\newcommand\mdate[3]{{#1}~{#2}~{#3}}  
\newcommand{\twom}{{\sc 2mass}}
\newcommand{\aspcs}{{ASP Conf.~Ser.}}
\newcommand{\sdbg}{{V1093\,Her}}
\newcommand{\sdbv}{{V361\,Hya}}
\newcommand\rel[2]{{{#1}\,=\,{#2}}}
\newcommand\temp[1]{\rel{\teff}{#1\,K}}
\newcommand\tempe[2]{\rel{\teff}{#1(#2)\,K}}
\newcommand\grav[1]{\rel{\logg}{#1}}
\newcommand\gravv[2]{\rel{\logg}{#1(#2)}}
\newcommand\helium[1]{\rel{\lheh}{#1}}
\newcommand\heliume[2]{\rel{\lheh}{\ensuremath{#1}(#2)}}
\newcommand\ellone{\rel{$\ell$}{1}}
\newcommand\ellmone{\rel{$\ell$}{--1}}
\newcommand\ellpone{\rel{$\ell$}{+1}}
\newcommand\elltwo{\rel{$\ell$}{2}}
\newcommand\emzero{\rel{$m$}{0}}
\newcommand\emone{\rel{$m$}{$\pm$1}}
\newcommand\emtwo{\rel{$m$}{$\pm$2}}
\newcommand\keplerack{
The authors gratefully acknowledge the \kep\ team and everybody who
has contributed to making this mission possible.
Funding for the \kepmi\ is provided by NASA's Science Mission Directorate.}
\newcommand\spectroack{
Spectroscopic observations were made with the William Herschel Telescope located at the
Observatorio del Roque de los Muchachos (ORM), the Nordic Optical Telescope also at the ORM, 
and the Steward Observatory 2.3-m Bok telescope.}
\newcommand\prosperityack{
This research has received funding from the European
Research Council under the European Community's Seventh Framework Programme
(FP7/2007--2013)/ERC grant N$^{\underline{\mathrm o}}$\,227224
({\sc prosperity}), as well as from the Research Council of KU~Leuven grant
GOA/2008/04.}

\newcommand{\target}{KIC\,1718290}
\newcommand{\ltarget}{SDSS\,J192300.68+371504.4}
\newcommand{\tmtarget}{\twom\,J19230068+3715044}

\shortauthors{\O stensen et al.}

\begin{document}

\title{KIC\,1718290: A helium-rich V1093-Her-like pulsator on the blue horizontal branch}
\author{
   R.~H.~\O stensen\altaffilmark{1},
   P.~Degroote\altaffilmark{1},
   J.~H.~Telting\altaffilmark{2},
   J.~Vos\altaffilmark{1},
   C.~Aerts\altaffilmark{1,3}, \\
   C.~S.~Jeffery\altaffilmark{4},
   E.~M.~Green\altaffilmark{5},
   M.~D.~Reed\altaffilmark{6}, and
   U.~Heber\altaffilmark{7}
}

\affil{$^1$ Instituut voor Sterrenkunde, K.U.~Leuven, Celestijnenlaan 200D, B-3001 Leuven, Belgium; \textcolor{blue}{roy@ster.kuleuven.be}}
\affil{$^2$ Nordic Optical Telescope, Apartado 474, 38700 Santa Cruz de La Palma, Spain}
\affil{$^3$ Department of Astrophysics, IMAPP, Radboud University Nijmegen, 6500 GL Nijmegen, The Netherlands}
\affil{$^4$ Armagh Observatory, College Hill, Armagh BT61 9DG, Northern Ireland}
\affil{$^5$ Steward Observatory, University of Arizona, 933 N.~Cherry Ave., Tucson, AZ 85721, USA}
\affil{$^6$ Department of Physics, Astronomy, and Materials Science, Missouri State University, Springfield, MO 65897, USA}
\affil{$^7$ Dr. Karl Remeis-Observatory \& ECAP, Astronomisches Inst., FAU Erlangen-Nuremberg, Sternwartstr.~7, 96049 Bamberg, Germany}

\begin{abstract}
   We introduce the first $g$-mode pulsator found to reside on the classical blue horizontal branch.
   One year of \kep\ observations of \target\ reveals a rich spectrum of low-amplitude modes with
   periods between one and twelve hours, most of which follow a regular spacing of \val{276.3}{s}.
   This mode structure strongly resembles that of the \sdbg\ pulsators, with only a slight
   shift towards longer periods. Our spectroscopy, however, reveals \target\ to be quite distinct
   from the sdB stars that show \sdbg\ pulsations, which all have surface gravities higher than
   \grav{5.1} and helium abundances depleted by at least an order of magnitude relative to the solar
   composition. We find that \target\ has \temp{22\,100}, \grav{4.72}, and a super-solar 
   helium abundance (\helium{--0.45}).
   This places it well above the extreme horizontal branch, and rather on
   the very blue end of the classical horizontal branch, where shell hydrogen burning
   is present. We conclude that \target\ must have suffered extreme mass loss during
   its first giant stage, but not sufficient to reach the extreme horizontal branch. 
\end{abstract}


\section{Introduction}

The \kep\ spacecraft is monitoring a 105 deg$^2$ field in the Cygnus--Lyrae
region, primarily to detect transiting planets \citep{borucki11}.
As a bycatch, pulsating stars are observed,
and these high-quality datasets are a treasure trove
for asteroseismology studies \citep{gilliland10a}.
In the first four quarters of the \kepmi\ a survey for pulsating stars
was made, and a total of 113 compact-pulsator
candidates were checked for variability \citep{ostensen10b,ostensen11b}.
The survey was extremely successful with respect to subdwarf-B (sdB) pulsators,
with discoveries including one clear \sdbv\ pulsator \citep{kawaler10a},
a total of thirteen \sdbg\ stars \citep{reed10a,kawaler10b,baran11b,2m1938}.

Preselection of \kep\ targets based on their photometric and
spectroscopic properties has provided us with the most outstanding
photometric lightcurves in the history of astronomy for a host of
known pulsator types.  In addition, the large archive of
\kep\ data\footnote{http://archive.stsci.edu/kepler/.},
mostly obtained for detecting exoplanets, provides
an inexhaustible source of spectacular lightcurves of all kinds of
variable stars, and unlimited opportunities for exciting new discoveries.
Here we present one such.
It was included in the color-selected sample of white-dwarf-pulsator candidates
that provided the only V777\,Her pulsator in the \kep\ field
\citep[KIC\,8626021][]{ostensen11c}.
However, the spectrum obtained in that survey placed it well outside of the known pulsator
ranges, and it was not considered further.
By chance, it was observed by {\em Kepler} as an exoplanet target, permitting serendipity
to trump our otherwise pinpoint-precision survey strategy.

Our spectrum shows that \target\ is technically a hot subdwarf of the sdB type, as it
appears as a B-type star with a rather high surface gravity, placing it
below the main sequence in the Hertzsprung-Russell diagram.
However, its surface gravity is lower and its helium abundance is much higher than
the majority of the sdBs, making it quite unusual.

After ascending the red-giant branch (RGB)
most stars with main-sequence masses less than \sval{\sim}{2}{\msol}
will undergo a core-helium flash and end up by the red clump,
where they spend \sval{\sim}{100}{Myr} burning
helium in the core, before ascending the asymptotic-giant branch (AGB).
If such stars suffer significant mass loss, then they will appear
bluer after helium ignition
while still having roughly the same luminosity. This produces the 
distinctive feature seen in color-magnitude diagrams of globular
clusters, known as the horizontal branch (HB).
HB stars possess a hydrogen-burning shell, whenever the
hydrogen envelope contains \sval{>}{0.02}{\msol}.
Stars with thinner envelopes will reside on
the extreme horizontal branch (EHB) \citep{heber86}.
Other HB stars are often designated as either red or blue, with
RHB stars encompassing the red clump and the notable class of RR\,Lyrae pulsators.
These are classical radial pulsators excited by the same heat mechanism acting in the
partial ionization zone of He$^+$/He$^{2+}$ that also drives
pulsations in Cepheids and $\delta$\,Scuti stars \citep{christy66}.
The BHB stars include all stars of spectral type A and B up to the start of the
EHB at \grav{5}.
The distribution of stars along the HB is a convoluted function of mass loss on the RGB, 
overall metallicity, and interactions with a close companion. 

\begin{figure*}[t!]$
\begin{array}{cc}
\parbox[b]{0.5\hsize}{
\includegraphics[width=0.97\hsize]{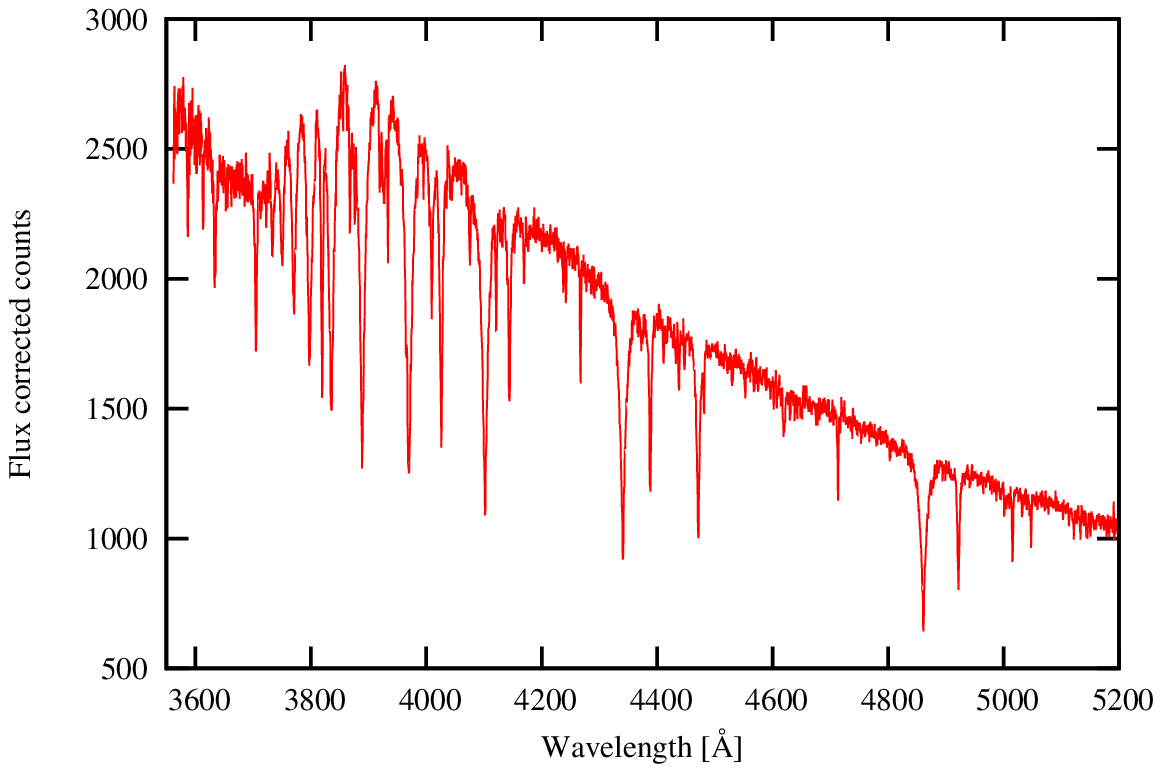} 
\includegraphics[width=0.94\hsize]{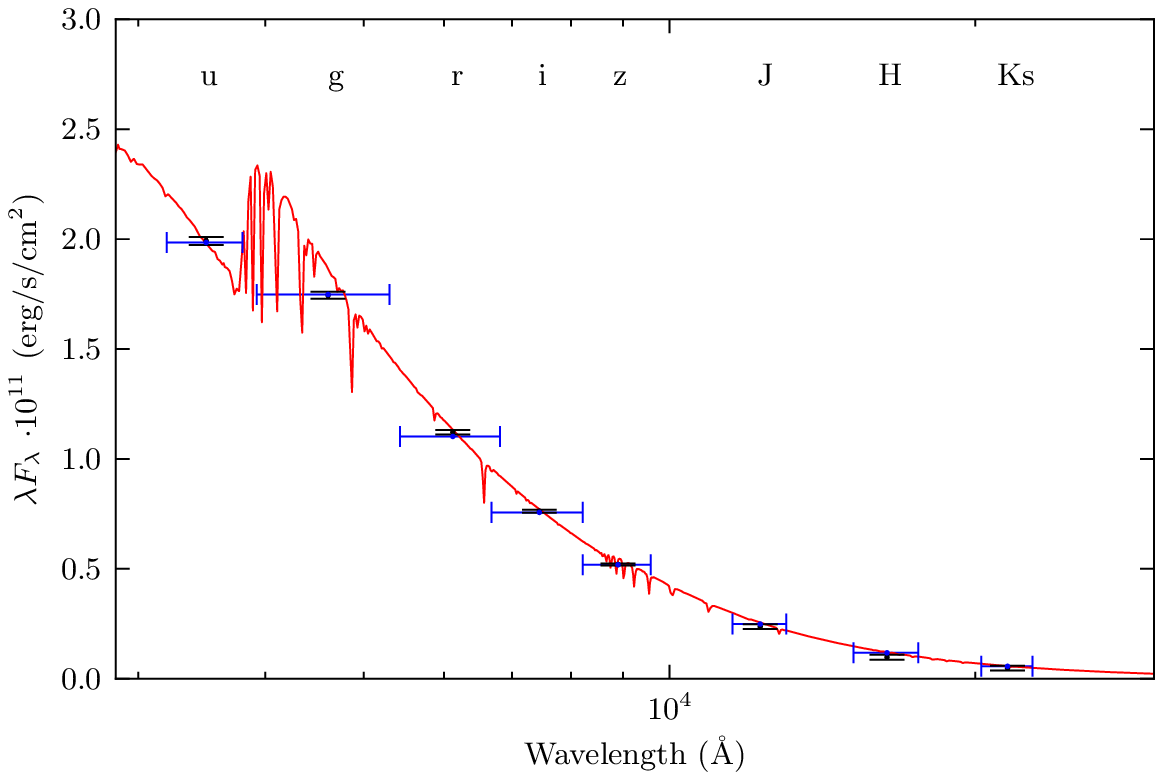}} &
\includegraphics[width=0.498\hsize]{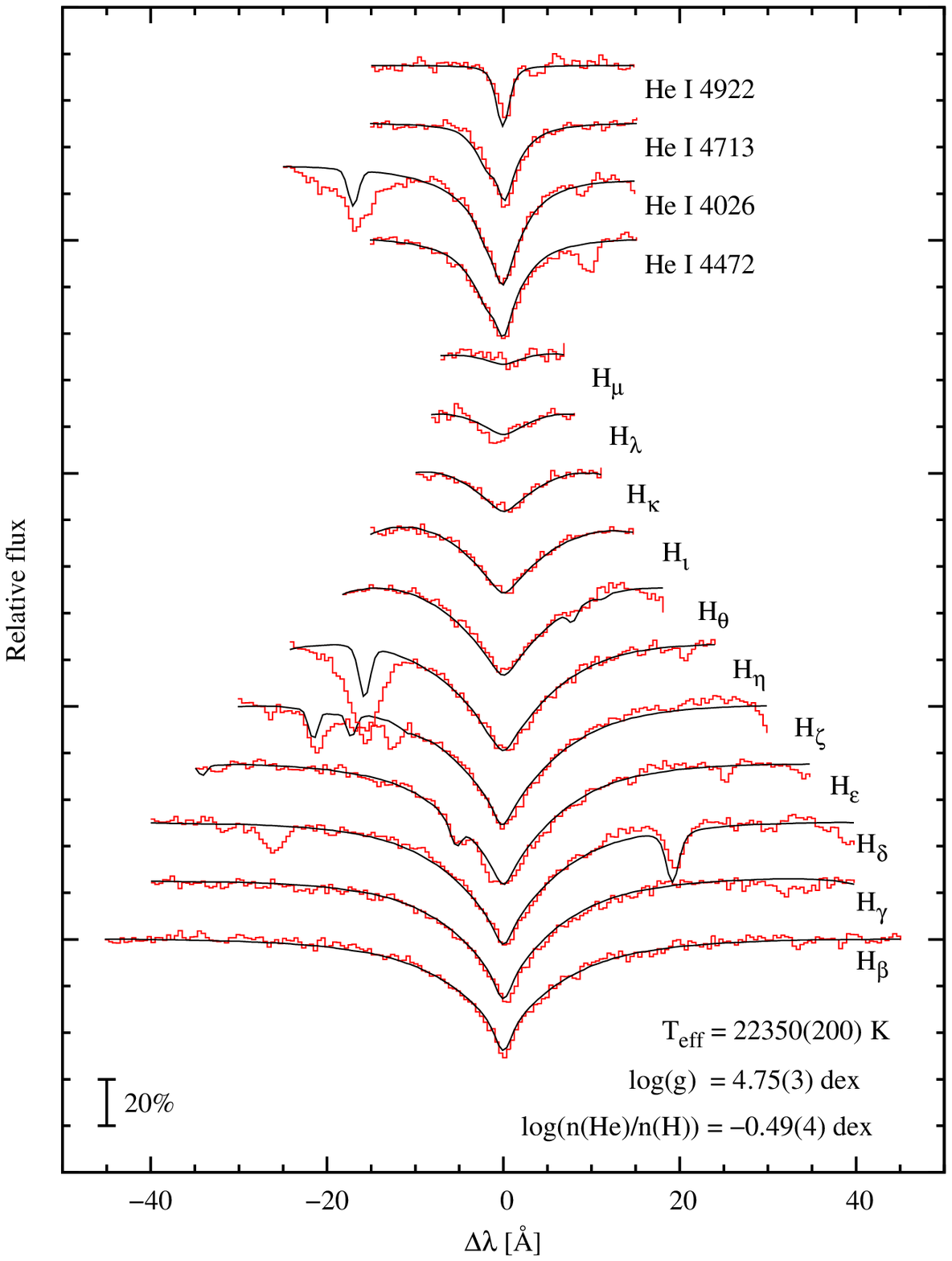}
\end{array}$
\put(-527,155){\bf a}
\put(-527,-8){\bf b}
\put(-260,155){\bf c}
\caption{
a:~The discovery spectrum of \target.
b:~SED fit to broad-band photometry.
c:~Line-profile fit to the WHT spectrum.}
\label{fig:sp}
\end{figure*}

The process that produces the EHB stars is thought to include a
large contribution from binaries that underwent mass transfer and
drastic mass loss in either common-envelope or Roche-lobe-overfilling
configurations on the RGB \citep{han02,han03}.
Whereas the majority of the sdB stars on the EHB are members of short-period
binaries \citep[][and references therein]{ostensen09}, such short-period binaries are
hard to find on the classical horizontal branch \citep{prsa08}.

Since the discovery of short-period pulsations
by \citet{kilkenny97}, the sdB stars have been extensively studied.
The $p$-mode pulsators are known as \sdbv\ stars
after the prototype, and constitute $\sim$10\%\ of sdBs
with temperatures between \sval{\sim}{28000}{K} and \val{36000}{K} \citep{sdbnot}. 
\citet{green03} discovered longer-period pulsations
in 75\%\ of stars between
\sval{\sim}{22000}{K} and \val{30000}{K}; the \sdbg\ stars.
Here we present the first
discovery of a star that exhibits long-period pulsations similar to the
\sdbg\ stars on the EHB, but is located on the BHB instead.

\section{Target properties}

\target\ (\rel{\mkep}{15.486})
was first noted to be a UV-excess star from SDSS photometry
\citep{SDSS}.
It appears as an exceedingly blue object\footnote{
http://cas.sdss.org/dr7/en/tools/chart/navi.asp?ra=290.753\&dec=37.251}
45'' to the east of BD+36$^\circ$3529.
The SDSS database lists it as \ltarget\ with
$ugriz=15.505(4)$, 15.293(3), $15.481(4),15.687(4),15.919(7)$.
The $u-g$ color term of +0.21 placed it above the red cutoff of our initial survey,
but in the 2010 search we allowed more generous margins
to catch possible ZZ-Ceti pulsators and sdBVs reddened by dust, as
is the case with \target. The target also appears as
\tmtarget\ with $JHK_s=15.50(5),$ $15.71(13),15.69(24)$.

We used the atmosphere models of \citet{kurucz93} to perform a
fit to the broadband photometry following the procedure in \citet{degroote11}.
The fit is consistent with a \tempe{21000}{1000},
and \rel{\ebv}{0.192(8)}, but provides no useful constraint on the surface gravity
(Fig.~\ref{fig:sp}b).
The optimal reddening agrees with \citet{schlegel98}, and we conclude that the rather
red color is caused by dust and not a composite-spectrum object.

We obtained a classification spectrum of \target\
with the {\sc isis} spectrograph on the 4.2-m William Herschel Telescope (WHT)
on \mdate{2010}{July}{3}, using the R600B grating (\rel{R}{1.7\,$\AA$ FWHM}),
giving S/N $\sim$45 per pixel.  
Over 28 days in March/April 2012 we obtained a further five 
S/N $\sim$50 spectra with the Nordic Optical Telescope
({\sc not}) using {\sc alfosc} with grism $\#$16 (\rel{R}{2.2\,$\AA$} FWHM).
The mean radial velocity (RV) is \sval{-}{29.2}{\kms}, with a standard deviation of
\val{7.5}{\kms}, demonstrating that low amplitude RV variability might be present.
The coverage is insufficient to rule out binarity at periods around
a month or longer.

\begin{figure*}[t!]
\begin{center}
\includegraphics[width=\hsize]{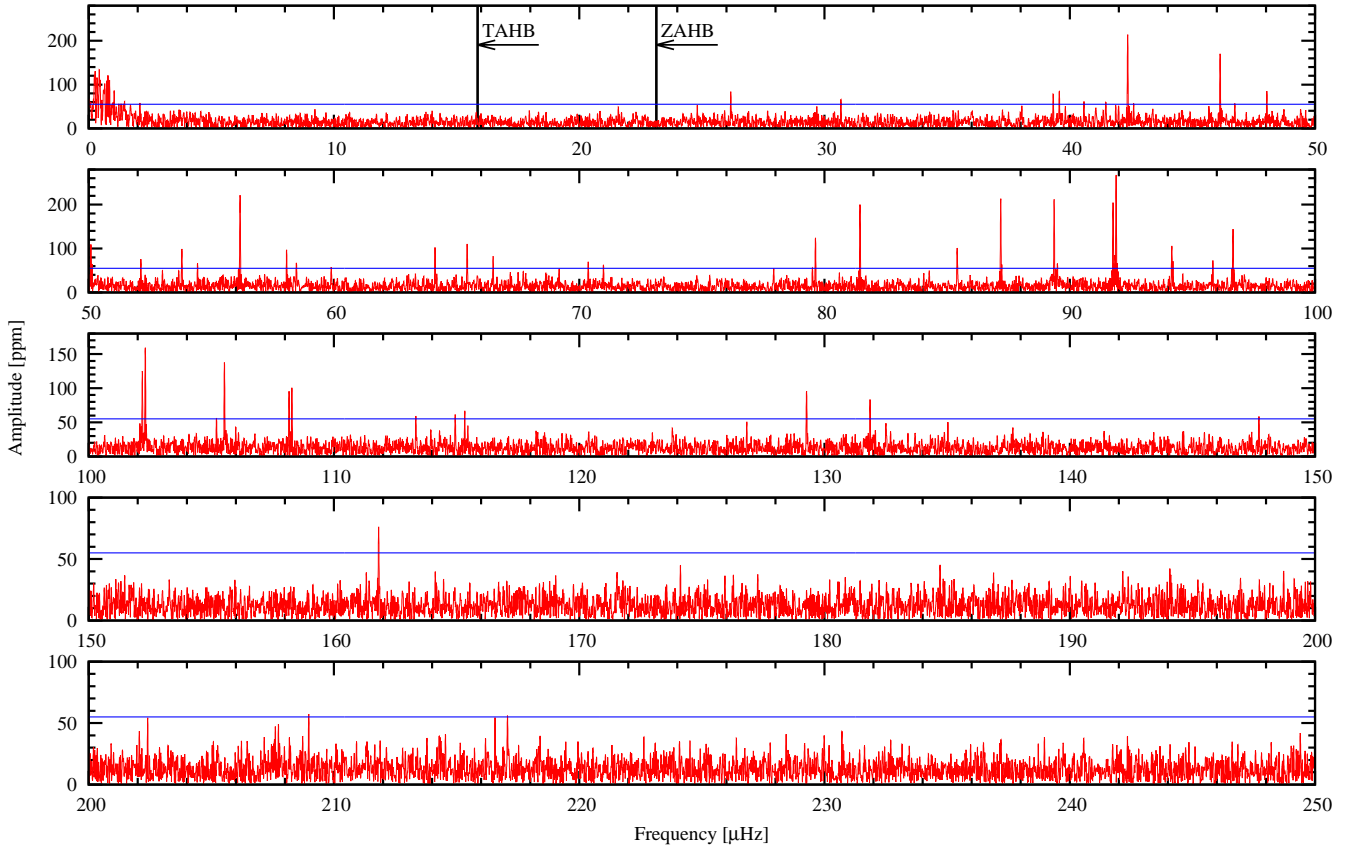}
\caption{
The FT of the \kep\ light curve of \target, including data from Q1--Q5.
The 4-$\sigft$ level is indicated with a horizontal line.
Vertical bars indicate the range for the theoretical lower limit for $g$-modes
on the ZAHB and TAHB, respectively.
}
\label{fig:kepft}
\end{center}
\end{figure*}

Standard spectral extraction and wavelength calibration was done using {\sc iraf}.
The spectrum shown in Fig.~\ref{fig:sp}a is the WHT observation after adjusting the slope
by an instrument response function derived from an observation of Feige\,67 obtained on
the same night. 

A fit to the spectra on an LTE grid suitable for these stars (see \citealt{heber00} and
\citealt{ramspeck01} for details), yielded an effective temperature of
\tempe{22350}{200}, a surface gravity of \gravv{4.75}{3}, 
and a helium abundance of \heliume{-0.49}{4} from the WHT spectrum (Fig.~\ref{fig:sp}c).
The mean {\sc not} spectrum yields \tempe{21796}{144}, \gravv{4.67}{3}, and \heliume{-0.40}{4} .

These parameters place \target\ well above the red, low-gravity edge of the EHB in
the \teff\--\logg\ plane, 
indicating that it could still have sufficient envelope mass to sustain a weak
hydrogen-burning shell.

\section{Frequency analysis}

We analysed the \kep\ photometry both from pipeline-extracted light curves and
pixel data. As the star is only $\sim$10 pixels away from the much brighter
star BD+36$^\circ$3529, some contamination is present. The \kep\ target catalog
lists it with contamination factors of 8.8, 11.6, 13.6 and 6.6\%\ depending on quarter.
We explored the background pixels surrounding the target in the pixel apertures to
see if any trace of a pulsation signature could originate from BD+36$^\circ$3529,
but found no significant signal in the frequency range where \target\ displays
its features.
After extracting the long-cadence light curves from quarter 1 through 5, we subtracted out the
slow trends introduced by the varying contamination from the neighbouring star with low-order 
polynomials and rapidly-decaying exponentials for the leading slope that often occurs
during the first days after the spacecraft has reacquired the field.
The light curve commences at BJD 2454964.51194, ends at BJD 2455371.16248, and spans
\val{406.6}{d}. The monthly gaps are typically about one day, with the largest gap being
five days.
The Fourier transform (FT) of this lightcurve is shown in Fig.~\ref{fig:kepft}.
A rich spectrum of low-amplitude peaks can be easily discerned, starting at
\sval{\sim}{25}{\uHz} and extending up to \val{217}{\uHz}.
The straight lines in Fig.~\ref{fig:kepft} indicate the 4-$\sigft$ level of \val{54.8}{ppm}.
We consider peaks of less than 4.2$\sigft$ or \val{58}{ppm}.
as tentative indications, as we have seen spurious detections at the 4.1-$\sigft$ level
in many other \kep\ targets \citep{ostensen10b}. 

The longest periods seen in \target\ are significantly longer than those seen
in the stars of \citet{reed11c}. So we computed the cutoff frequency for
$g$-modes according to the prescription of \citet{hansen85} and found that the longest
period that can be sustained is \sval{\sim}{0.5}{d} at the zero-age HB and peaks at \sval{\sim}{0.73}{d} close
to the terminal-age HB, when using the \val{0.476}{\msol} model described in the next section.
In Fig.~\ref{fig:kepft} the cutoff-frequency range is marked with vertical bars.

\begin{deluxetable}{lrrrrlll}
\tablecaption{List of frequencies \label{tbl:peaks}}
\tablewidth{0pt}
\tablehead{
\colhead{ID} & \colhead{Freq.} & \colhead{$P$} & \colhead{$A$} & 
\colhead{Phase} & \colhead{$n_{\ell=1}$} & \colhead{$n_{\ell=2}$} & 
\colhead{Note} \\
\colhead{} & \colhead{(\uHz)} & \colhead{(d)} & \colhead{(ppm)} & 
\colhead{(cyc)} & \colhead{} & \colhead{} & \colhead{}
}
\startdata
$f_{1}$   & 24.81 & 0.4665 & 54 & 0.290 & & & t \\
$f_{2}$   & 26.18 & 0.4421 & 84 & 0.030 & & & u \\
$f_{3}$   & 30.67 & 0.3774 & 67 & 0.503 & & & u \\
$f_{4}$   & 39.32 & 0.2943 & 79 & 0.812 & 91\dag & & d \\
$f_{5}$   & 39.57 & 0.2925 & 82 & 0.864 & 91 & & d \\
$f_{6}$   & 40.58 & 0.2852 & 58 & 0.588 & 89\dag & & t \\
$f_{7 }$  & 41.47 & 0.2791 & 56 & 0.758 & 87\dag & 151 & t \\
$f_{8 }$  & 41.88 & 0.2764 & 57 & 0.290 & 86 & & \\
$f_{9 }$  & 42.36 & 0.2732 &214 & 0.635 & 85 & & \\
$f_{10}$  & 46.13 & 0.2509 &171 & 0.357 & 78 & & \\
$f_{11}$  & 46.73 & 0.2477 & 57 & 0.768 & 77 & 134 & \\
$f_{12}$  & 48.04 & 0.2409 & 88 & 0.980 & 75 & & \\
$f_{13}$  & 50.09 & 0.2311 &111 & 0.939 &    & 125 & \\
$f_{14}$  & 52.13 & 0.2220 & 74 & 0.381 & 69 & 120 & \\
$f_{15}$  & 53.80 & 0.2151 &101 & 0.539 & 67\dag & & \\
$f_{16}$  & 54.44 & 0.2126 & 70 & 0.504 & 66 & 115 & p \\
$f_{17}$  & 56.17 & 0.2061 &222 & 0.248 & 64 & & \\
$f_{18}$  & 58.07 & 0.1993 & 96 & 0.321 & 62\dag & & \\
$f_{19}$  & 58.47 & 0.1980 & 68 & 0.611 &    & 107 & \\
$f_{20}$  & 59.89 & 0.1933 & 58 & 0.333 & 60 & & t \\
$f_{21}$  & 64.12 & 0.1805 &103 & 0.099 & 56 & & c\\
$f_{22}$  & 65.43 & 0.1769 &111 & 0.309 & 55\dag & & \\
$f_{23}$  & 66.49 & 0.1741 & 80 & 0.102 & 54 & 94 & p \\
$f_{24}$  & 70.36 & 0.1645 & 70 & 0.888 & 51 &  & \\
$f_{25}$  & 70.99 & 0.1630 & 61 & 0.639 &    & 88 & \\
$f_{26}$  & 77.93 & 0.1485 & 56 & 0.462 & 46 & & t \\
$f_{27}$  & 79.51 & 0.1456 & 55 & 0.159 & 45 & & t, d, p \\
$f_{28}$  & 79.63 & 0.1454 &125 & 0.364 & 45 & & d \\
$f_{29}$  & 81.45 & 0.1421 &201 & 0.026 & 44 & &c\\
$f_{30}$  & 85.41 & 0.1355 &100 & 0.097 & 42 & 73 & \\
$f_{31}$  & 87.19 & 0.1328 &214 & 0.879 & 41\dag & & \\
$f_{32}$  & 89.36 & 0.1295 &213 & 0.474 & 40 & & d \\
$f_{33}$  & 89.49 & 0.1293 & 73 & 0.639 & 40 & & c, d\\
$f_{34}$  & 91.77 & 0.1261 &205 & 0.994 & 39 & 68 & d \\
$f_{35}$  & 91.89 & 0.1260 &268 & 0.129 & 39 & & d \\
$f_{36}$  & 94.16 & 0.1229 &103 & 0.805 & 38 & & c, d\\
$f_{37}$  & 94.21 & 0.1229 & 57 & 0.955 & 38 & & t, d \\
$f_{38}$  & 95.84 & 0.1208 & 74 & 0.613 &    & 65 & p \\
$f_{39}$  & 96.66 & 0.1198 &144 & 0.879 & 37 & & \\
$f_{40}$  &102.19 & 0.1133 &125 & 0.083 & 35 & & d \\
$f_{41}$  &102.31 & 0.1131 &160 & 0.521 & 35 & & c, d\\
$f_{42}$  &105.53 & 0.1097 &138 & 0.236 & 34\dag & 59 & \\
$f_{43}$  &108.17 & 0.1070 & 96 & 0.181 & 33 & & c, d\\
$f_{44}$  &108.28 & 0.1069 &101 & 0.294 & 33 & & c, d\\
$f_{45}$  &113.34 & 0.1021 & 58 & 0.183 &    & 55\dag & c, t \\
$f_{46}$  &114.94 & 0.1007 & 61 & 0.919 & 31 & 54\dag & d \\
$f_{47}$  &115.33 & 0.1004 & 68 & 0.855 & 31 & 54 & d \\
$f_{48}$  &129.27 & 0.0895 & 96 & 0.749 &    & 48 & \\
$f_{49}$  &131.86 & 0.0878 & 83 & 0.770 & 27 & & c, p \\
$f_{50}$  &147.71 & 0.0784 & 58 & 0.768 & 24 & 42 & c, p \\
$f_{51}$  &161.82 & 0.0715 & 76 & 0.954 & 22 & &  \\
$f_{52}$  &208.97 & 0.0554 & 57 & 0.323 & 17\dag & & t, p  \\
$f_{53}$  &216.57 & 0.0534 & 54 & 0.084 &    & & c, t, p \\
$f_{54}$  &217.08 & 0.0533 & 55 & 0.478 &    & & t, p \enddata 
\tablecomments{
\dag: Uncertain ID ($>$10\%\ off sequence), \\
c: Possible combination frequency, \\
d: Doublet component, \\
p: Non-prewhitenable peak,  \\
t: Tentative mode (between 4 and 4.2$\sigft$), \\
u: Unidentifiable peak}
\end{deluxetable}

\begin{figure}[t!]
\centering
\includegraphics[width=\hsize]{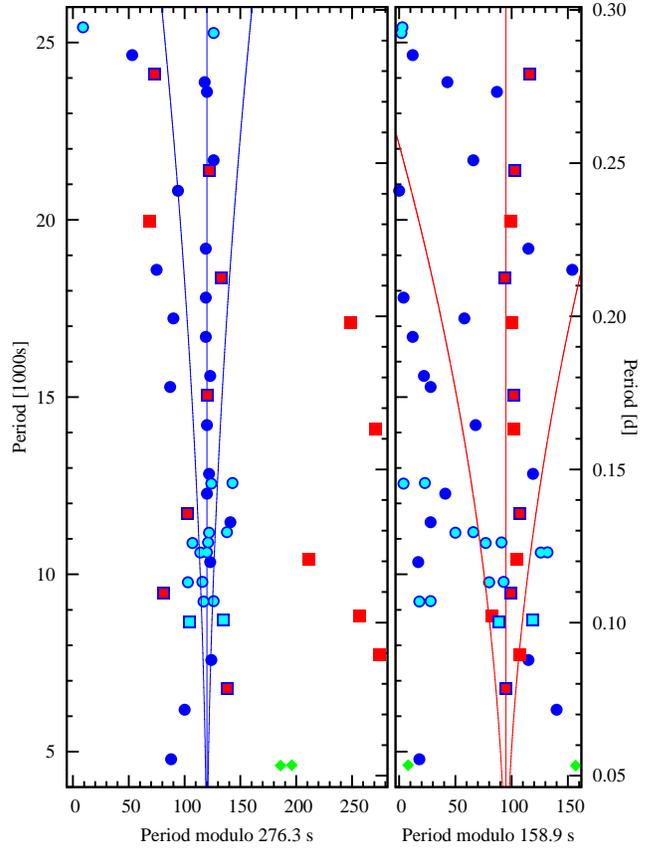}
\caption{
\'{E}chelle diagram for \target.
Bullets indicate \ellone\ modes (blue in the on-line version) and boxes \elltwo\ modes (red).
Periods that may belong to either sequence are marked with outline boxes (red core with blue outline).
Periods which are part of doublets are marked with hollow bullets or boxes (blue with cyan cores),
for \ellone\ or \elltwo\ modes respectively.
Two points above \val{26000}{s} (not shown)
and two points below \val{5000}{s} (green diamonds) do not fit either sequence.
The curves indicate the width of an \ellone\ triplet (left panel)
and of an \elltwo, \rel{$m$}{--2,0,+2} triplet (right panel), in the absence of any
trapping.
}
\label{fig:echelle}
\end{figure}

The frequencies and periods of the detected peaks are listed in Table~\ref{tbl:peaks},
together with their amplitudes and phases relative to the first data point.
The resolution of the dataset is \sval{\sim}{0.03}{\uHz}, and can
be taken as the error on the measured frequencies.
We detected 54 frequencies, of which eight are close pairs.
No triplets or higher-order multiplets are seen.
All of the frequencies except nine 
can be prewhitened out of the light curve, demonstrating that they have stable amplitudes.
A few of the peaks may also be combination frequencies. 

The most discernible feature of this list is the regular sequence of peaks with an even
period spacing of \rel{$\Delta\Pi_1$}{276.3\,s}. This is slightly longer than the
longest spacing detected in the 13 \sdbg\ stars observed by \kep\ and studied by
\citet{reed11c,reed12}, of \val{271.15}{s}, and associated with \ellone\ modes.
For all but two of those stars a second sequence was found to correspond to
\elltwo\ modes with a spacing 1/$\sqrt{3}$ times that of the \ellone\
sequence, exactly as predicted by theory. But whereas the \ellone\ sequence in
\target\ is clearer than in any of these 13 \sdbg\ pulsators, the \elltwo\
sequence which should be at \rel{$\Delta\Pi_2$}{159.5} is tentative at best.
While an unusually large number of modes fall on the \val{276.3}{s} sequence within
$\pm$1\% (14/54), a much larger number of modes fits within a more generous
$\pm$10\%\ interval (35/54).
Some of the modes appear to be doublets, including some of the strongest modes such
as $f_{34,35}$.
The same frequency splitting is seen in the pairs
$f_{27,28}$, $f_{32,33}$, $f_{34,35}$, $f_{36,37}$, $f_{40,41}$ and $f_{43,44}$
(splittings between 0.115 and \val{0.125}{\uHz}).
The  $f_{4,5}$ doublet has twice that (\val{0.25}{\uHz}) and the $f_{46,47}$ doublet
has three times the splitting (\val{0.39}{\uHz}).
As the majority of these doublets fit the \ellone\ period spacing sequence,
a possible interpretation is that they are the \emone\ components of \ellone\ triplets.
Unless one can invoke some other mechanism that would systematically suppress
either the \ellmone\ or the \ellpone\ component so that one never sees any triplets,
the most plausible explanation is that the viewing angle is aligned with the pulsation equator,
where the \emzero\ components suffer large cancellation.
Following this interpretation, the \elltwo\ modes which suffer the least geometric
cancellation at this inclination are the \emtwo\ components.
Using \rel{$C_{n1}$}{0.5} and \rel{$C_{n2}$}{0.16}, as indicated by the asymptotic theory,
the \elltwo, \emtwo\ components
would be split by \sval{\sim}{0.40}{\uHz} which fits the $f_{46,47}$ doublet.
If we accept this argument and take \val{0.12}{\uHz} to be the rotationally induced
frequency splitting for an \ellone, \emone\ pair, then
the rotation period is \sval{\sim}{100}{d}. A similar rotationally induced splitting was found
for the sdBV+WD binary KIC\,11558725 \citep[\val{0.13}{\uHz},][]{telting12b}.
Note that 
the \ellone, \emone\ modes will impose a scatter in the
\'{e}chelle diagram that increases with increasing period, as indicated by the curves in
Fig.~\ref{fig:echelle}.  For the $f_{4,5}$ doublet the 
splitting is \sval{\sim}{100}{s}, making secure identification
by sequence spacing rather hopeless. As can be seen in the left panel of
Fig.~\ref{fig:echelle}, only five modes cannot be fitted on a generous \ellone\ sequence,
when tentative peaks and the long-period modes $f_{2,3}$ are ignored.
If we impose a sequence at \rel{$\Delta\Pi_2$}{158.9\,s}, the remaining five modes can
also be explained, and we see that seven modes fit both sequences. The \elltwo\
sequence is far from convincing, but forthcoming \kep\ data should reduce the noise
and may substantiate it.
We performed linear-regression fitting to the identified periods and a
Kolmogorov-Smirnov (KS) test as in \citet{reed11c}.
The fit to the period spacings were 276.3(1) and 158.91(5) seconds with
KS significances of 6.2 and 0.98$\sigma$
for the \ellone\ and \elltwo\ sequences, respectively. 

\begin{figure}[t!]
\includegraphics[width=\hsize]{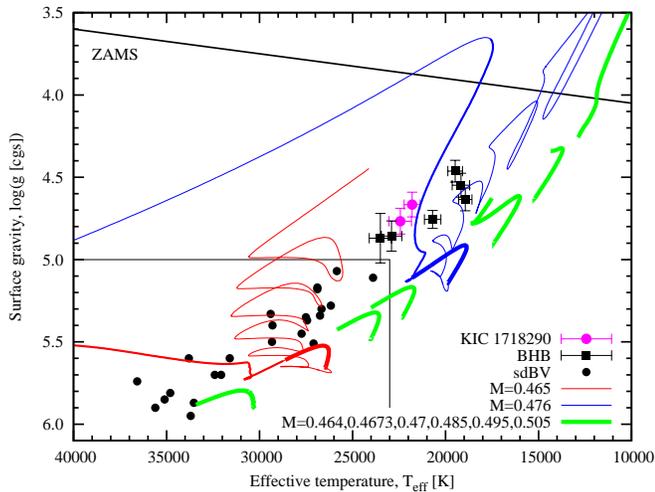}
\caption{The \teff\--\logg\ plane with evolutionary models and some sample stars.
The target and six other intermediate-helium stars
are plotted with error-bars indicating three times the formal-fitting errors.
The parameters from the WHT and the NOT spectrum are shown as bullets
with error bars. The classical EHB region is outlined, and encompasses
a number of sdB pulsators that we have fitted on the same model grid
as \target. }
\label{fig:tgplot}
\end{figure}

\section{Discussion}

The pulsation spectrum of \target\ shows the same richness and characteristic
period spacings as that
of the thirteen \sdbg\ pulsators monitored by \kep, with the exception that the
pulsations have longer periods (by about a factor of two).

We constructed a MESA model for the Sun \citep{paxton11} and
evolved it from the pre-main-sequence phase up to the point of core-helium ignition,
assuming standard Reimers mass loss \citep[][with \rel{$\eta$}{0.5}]{reimers75}.
Just before helium ignition we stripped the stellar model of most of its envelope,
leaving only a residual hydrogen envelope on top of the He-core.
The model was then relaxed to a new equilibrium state.
Different models with a varying thickness of the H-layer were then evolved
through the He-flashes, the He-core-burning and He-shell-burning stage.
Note that these models were computed with default MESA parameters, and do not
reproduce the position of the observed EHB very well. A number of model parameters
can be tweaked to change the position of the EHB. For instance, it
is very sensitive to the exact mass where the helium flash occurs.
As shown by \citet{dorman93}, 
increasing the helium abundance and/or reducing the metallicity would shift it in
the right direction, but since we have no reason to assume unconventional initial
abundances that apply to all progenitors of BHB/EHB stars, we decided to leave
the offset as it is for now.

In Fig.~\ref{fig:tgplot} we show two extended tracks from this set, and six tracks
truncated to show just the core-burning stage where the stars
spend most of their time.
The short-lived stages prior to stable core-helium burning and after the end
of shell-helium burning are shown with thin lines. The \sval{\sim}{80}{Myr} core-helium
burning stages are plotted with thick lines, and the \sval{\sim}{62}{Myr} shell-helium burning
stages are plotted with intermediately-thick lines. The high-gravity evolutionary sequence
(\rel{M}{0.465\,\msol}) is typical for an EHB star, while the lower-gravity sequence
(\rel{M}{0.476\,\msol})
is more representative of the observed position of \target. EHB stars evolve directly
to the white-dwarf cooling curve after ending their helium burning, while normal HB stars
ascend the AGB. The \rel{M}{0.476\,\msol} model displays an intermediate
behaviour, where its radius expands substantially and the temperature drops slightly,
but still never getting close to the AGB; just barely reaching the \teff/\logg-level
of the main sequence before faltering and starting the contraction that will eventually
take it to the white-dwarf cooling curve.

In Fig.~\ref{fig:tgplot} the location of target and six other intermediate-helium stars,
PG\,0229+064, PG\,0848+186, PG\,1400+389, PG\,2356+167, Balloon 82000001 and CPD--20$^\circ$1123
are shown.
CPD--20$^\circ$1123 was recently shown to be a binary with an orbital period of \val{2.3}{d}
\citep{naslim12}. For the other stars, no information about binarity has been published.
All these stars have \lheh\ between --1.0 and --0.5, i.e.~they all have somewhat super-solar
helium abundances. Most stars classified as He-sdB and He-sdO
are extremely helium rich; their photospheres have more helium than hydrogen by number,
sometimes by a factor $10^3$, and may be associated with double-helium-white-dwarf
mergers \citep{zhang12}. The intermediate-helium sdB stars appear confined to
a more narrow region of the \teff/\logg\ plane than the extreme-helium sdBs are,
implying a more narrow core-mass range.

At higher gravities in Fig.~\ref{fig:tgplot}, the pulsating stars from the \kep\ sample are
plotted with their physical parameters from \citet{ostensen10b,ostensen11b},
and further down the EHB we include some \sdbv\ pulsators from \citet{sdbnot} that were fitted
on the same LTE grid as the other stars. 
Note that the number of stars marked in Fig.~\ref{fig:tgplot} is in no way representative
of the population densities of these stars. The EHB stars in the figure are just 24
out of more than 2000 known sdBs \citep{ostensen06}, while the BHB stars indicated
include the majority of stars similar to \target\ described in the literature.
Thus, as pointed out by \citet{saffer94,saffer97},
what is known as the second Newell gap from studies of globular clusters \citep{newell76}
is neither a forbidden region on the BHB nor due to selection effects, but rather a very
underpopulated region that coincides with the position where the envelope has the
critical mass required for hydrogen-shell burning. Explaining the exact process that
produces the gap remains a challenge.
As one proceeds from this gap towards the redder part of the HB it becomes exceedingly
hard to distinguish between field-horizontal-branch-B stars and main-sequence-B stars, but
high-resolution spectroscopy can distinguish their peculiar chemical composition and
low rotation rates \citep{hambly97}.  
\section{Conclusions}

We have shown that \target\ is spectroscopically an intermediate-helium-sdB
star, most likely associated with the BHB. Photometric
data of extremely high precision and coverage obtained with \kep\ have revealed
a pulsation spectrum with a clear period spacing of \val{276.3}{s},
indicative of a core-helium burning star similar to the \sdbg\ pulsators.
There is no reason to believe that the driving mechanism responsible for exciting the
pulsations in \target\ is any other than the same $\kappa$-mechanism that drives
the pulsations in the \sdbg\ and \sdbv\ stars \citep{fontaine03}. The red edge of
the \sdbg\ instability strip has never been firmly established, due to the sharp
drop in the number of observed EHB stars at around \val{24\,000}{K}.
\sdbg\ pulsations in stars at the blue tip of the BHB should not really come
as a big surprise, as early models for these stars predicted pulsations
at the cool end of the EHB and beyond \citep{jeffery06a}.

\target\ must have suffered extreme mass loss during its evolution in order to reach
its present configuration. While our SED fit displays no sign of any main-sequence companion,
a white-dwarf companion is not excluded. The \kep\ photometry
rules out any close object, since a period of \val{10}{d} or less would be readily detected
from Doppler beaming in the light curve, unless the orbit is seen face on \citep{telting12b}.
White-dwarf companions with orbits \sval{\sim}{30}{d} are
harder to detect with \kep\ due to the monthly data-download cycle, and the
reacquisition of the field after each such manoeuvre introduces trends in the data that must
be filtered out. Spectroscopic follow-up observations are being undertaken, but have not
yet provided any stronger limits than can be determined from photometry.

We conclude that we have observed $g$-mode pulsations in a star at the blue
end of the classical horizontal branch that are driven by the same $\kappa$-mechanism
as in the \sdbg\ pulsators and slowly-pulsating main-sequence-B stars. 
The pulsations are of longer period and lower maximum amplitude
than has been observed in any of the \kep\ \sdbg\ targets, making this type of
pulsator practically undetectable with ground-based observations.
In analogy with the naming sdBV for the group consisting of \sdbv\ and \sdbg\ pulsators,
we propose to call \target\ and similar BHB variables for BHBV stars.
The rarity of BHB stars at similar temperatures will make it hard to find other
examples of such pulsators in the limited field-of-view of \kep. However, the question
of how far towards the red along the BHB $g$-mode pulsations may extend remains a
question to be explored.

\acknowledgements
\keplerack
\spectroack
\prosperityack

\end{document}